\documentclass[pra,twocolumn,showpacs,preprintnumbers,amsmath,amssymb]{revtex4}
\usepackage{mathrsfs}
\usepackage{bbm}
\usepackage{amsfonts}
\usepackage{tipa}
\usepackage{CJK}
\usepackage{epsfig,graphicx}
\usepackage{amstext}
\usepackage{amsmath}
\usepackage{graphicx}
 \begin{document}
 %%%%%%%%%%%%%%%%%%%%%%%%%%%%%%%%%%%%%%%%%%%%%%%%%%%%%%%%%%%%%%%%%%%%%%
 %TCIDATA{OutputFilter=Latex.dll}
 %TCIDATA{Version=5.00.0.2552}
 %TCIDATA{<META NAME="SaveForMode" CONTENT="1">}
 %TCIDATA{LastRevised=Wednesday, June 22, 2005 16:21:09}
 %TCIDATA{<META NAME="GraphicsSave" CONTENT="32">}

 \title{Tuning quantum discord in Josephson charge qubits system}
 \author{Bo-Xing Shang$^{1,}$}
 \email{shangboxing@126.com}
 \author{Xiao-Qiang Xi$^{2,}$}
 \email{xxq@xupt.edu.cn}
 \affiliation{$^{1}$School of Science, Xi'an University of Posts and Telecommunications, Xi'an 710121, China \\
 $^{2}$Institute of Internet of Things and IT-based Industrialization,
 Xi'an University of Posts and Telecommunications, Xi'an 710121, China}

\begin{abstract}
A type of two qubits Josephson charge system is constructed in this
paper, and properties of the quantum discord (QD) as well as the
differences between thermal QD and thermal entanglement were
investigated. A detailed calculation shows that the magnetic flux
$\Phi_{Xk}$ is more efficient than the voltage $V_{Xi}$ in tuning
QD. By choosing proper system parameters, one can realize the
maximum QD in our two qubits Josephson charge system.
 \end{abstract}


\maketitle \section{Introduction}\label{sec:1}

Quantum computation can process information with efficiency that
cannot be achieved in a classical way. The key reason for this high
efficiency is the existence of quantum correlations in the
computational system. As a typical quantum correlation measure,
entanglement has been extensively studied in the past two decades
\cite{RMP-entangle}. However, when mixed states are taken into
account, the role of entanglement turns to be less clear in certain
quantum tasks \cite{aop2332,qt1,qt2}. Particularly, in the protocol
of deterministic quantum computation with one qubit
\cite{dqc1,dqc2}, the estimation of the normalized trace of a
unitary matrix can be attained in a number of trials that do not
scale exponentially with its dimension. As quantum discord (QD),
\cite{PRL_88_017901,JPA_34_6899} other than entanglement, is present
in the final stage of the aforementioned task, it has been determined to be another
resource that is essential for quantum computation
\cite{PRL_100_050502}.

Due to the importance of quantum information processing (QIP), QD
has recently been the research focuses of scientists. The corresponding
investigation includes its quantification \cite{Modi,Giorgi,luosl,
min}; its relation with uncertainty principle \cite{uncer1,
uncer2,uncer3}; and other related issue (See Ref. \cite{RMP-modi}
for an overview). Physically, there are many systems that can be used
to realize QD, such as the spin-chain \cite{PRA_81_044101,
PRA_81_032120,zhang}, the atomic \cite{ncxu,aop1,aop2,
NJP_12_073009}, the spin-boson \cite{PRA_81_064103}, and the NMR
systems \cite{PRA_81_062181}. Recent studies also provided evidence
that QD is a resource in the tasks of entanglement distribution
\cite{PRL_109_070501}, remote state preparation
\cite{NatPhys_8_666}, and information encoding \cite{NatPhys_8_671}.
Experimentally accessible measures of QD have also been proposed
\cite{PRA_84_032122}. The superconducting qubits have been
considered as possible candidates in various QIP tasks
\cite{Nature_453_1031, Nature_449_443, PRL_89_197902, PRA_74_052321,
PRL_95_087001}. It has been experimentally demonstrated that they
posses macroscopic quantum coherence and can be used to construct
the conditional two-qubit gate. It is then necessary to scale upwords
to many qubits to perform the complex QIP tasks. In reference
\cite{PRA_74_052321}, Liu \emph{et al} proposed to use a
controllable time-dependent electromagnetic field to couple a
superconducting qubit with the data bus, where the quantum
information can be transferred from one qubit to another.

In this paper, we introduce a two qubits Josephson charge system
\cite{PRL_89_197902} to disclose the dependence of the thermal QD on
temperature $T$ and inter-qubit coupling strength $J_{ij}$ that
is controlled by the external flux $\Phi_{e}$ and local fluxes
$\Phi_{X_{i,j}}$, as well as $\varepsilon_{i}(V_{Xi})$ that is
controlled by the gate voltage $V_{Xi}$. We want to find the proper
system parameters that can make QD to realize its maximum value. We will
also compare QD with entanglement of formation (EoF) \cite{eof} and
reveal their differences.

The paper is organized as follows. Sec. \ref{sec:2} is a review of
the definitions of QD and EoF for the bipartite state; Sec.
\ref{sec:3} includes the introduction of the model, and the explicit
methods for tuning QD. Sec. \ref{sec:4} is a short summarization.

\section{Measures of quantum correlations}\label{sec:2}

One of the basic problems in QIP is to find the robustness essence
of the quantum correlations in a composite system. With this
motivation, we study the tuning of QD in a two qubits Josephson
charge system, and compare its behavior with that of the
entanglement measure EoF.

QD, as a measure of non-classical correlation, is defined as the
discrepancy between quantum mutual information and the classical
aspect of correlation, which can be defined as the maximum
information of one subsystem that can be obtained by performing a
measurement on the other subsystem. If we restrict ourselves to the
projective measurements performed locally on a subsystem
described by a complete set of orthogonal projectors $\{\Pi_{k}\}$,
then the quantum state will change as
$\rho_{b|k}=(\Pi_{k}\otimes I)\rho(\Pi_{k}\otimes I)/p_{k}$, where
$ I$ is the identity operator for subsystem $b$, and $p_{k}={\rm
tr}[(\Pi_{k}\otimes I)\rho(\Pi_{k}\otimes I)]$ is the probability
for obtaining the measurement outcome $k$ on $a$. The classical
correlation can be obtained by maximizing
$J(\rho|\{\Pi_{k}\})=S(\rho^{b})-S(\rho|\{\Pi_{k}\})$ over all
$\{\Pi_{k}\}$, where $S(\rho|\{\Pi_{k}\})=\sum_{k}p_{k}
S(\rho_{b|k})$ is a generalization of the classical conditional
entropy of the subsystem $b$. Explicitly, QD is defined as the
minimum difference between $I(\rho)$ and $J(\rho|\{\Pi_{k}\})$ as
 \begin{eqnarray}
  D=I(\rho)-\max_{\{\Pi_{k}\}}J(\rho|\{\Pi_{k}\}),
 \end{eqnarray}
where the maximum is taken over by the complete set of $\{\Pi_{k}\}$.
The intuitive meaning of QD may be interpreted as the minimal loss
of correlations due to measurement. It disappears in states with only
classical correlation and survives in states with quantum
correlation.

The EoF for a two-qubit state can be derived as \cite{eof}
 \begin{equation}
 E=H\left(\frac{1+\sqrt{1-C^{2}}}{2}\right),
 \end{equation}
where $H(\tau)=-\tau\log_{2}\tau-(1-\tau)\log_{2}(1-\tau)$ is the
binary Shannon entropy; $C=\max\{0,\lambda_{1}-\lambda_{2}
-\lambda_{3}-\lambda_{4}\}$ is the time-dependent concurrence
\cite{conc} with $\lambda_{i}~(i=1,2,3,4)$ being the square roots
of the eigenvalues of $R=\rho(\sigma_{y} \otimes\sigma_{y})
\rho^{*}(\sigma_{y}\otimes \sigma_{y})$ arranged in decreasing
order; $\rho^{*}$ is the complex conjugation of $\rho$ in the
standard basis; and $\sigma_{y}$ is the second Pauli matrix.

\section{Josephson charge-qubit system}\label{sec:3}

We first introduce the proposed Josephson charge-qubit system
\cite{PRL_89_197902}, which consists $N$ Cooper-pair boxes that are
coupled by a common superconducting inductance $L$ (see Fig. 1).
Each Cooper-pair box is weakly coupled by two symmetric dc
Superconducting Quantum Interference Device (SQUIDs) and biased by an
applied voltage $V_{Xk}$ through a gate capacitance $C_{k}$. The two
charges qubit system can be achieved by adjusting the voltage
$V_{Xk}$, which can control the number of Cooper-pair box within the
island. On the other hand, the Josephson coupling energy can be
adjusted by controlling the magnetic flux $\Phi_{Xk}$ through the
two SQUID loops of the $k$-th Cooper-pair box. It should be noted
that in Fig. 1 we considered only the nearest neighbor coupling
energy between charge qubits and ignored self inductance of the
SQUID loop and the electrical inductance of the superconducting wire
that connect the two charge qubits. If the interactions between the
two charge qubits are not the nearest neighbor, we can not ignore
electrical inductance of the superconducting wire. In this case, the
form of the system Hamiltonian remains unchanged, but their intrabit
coupling $\bar{E}_{J_i}$ and $J_{ij}$ may be changed.
 %%%%%%%%%%%%%%%%%%%%%%%%%%%%%%%%%%%%%%%%%%%%%%%%%%%%%%%%%%%%%%%
 \begin{figure}
 \centering \resizebox{0.4\textwidth}{!}{%
 \includegraphics{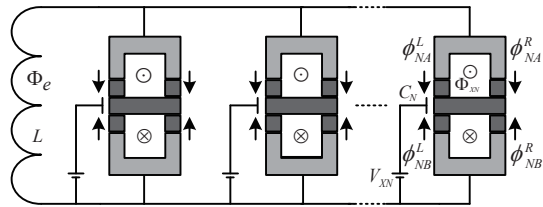}}
 \caption{Schematic diagram of the Josephson charge-qubits system.
 Here $\Phi_{e}$ and $\Phi_{XN}$ represent respectively the magnetic
 flux crossing electrical inductance $L$ and $N$th SQUID,
 while $\Phi_{NA(B)}^{L(R)}$ represents the phase Josephson
 junction. $V_{XN}$ is the voltage through a gate capacitance $C_{N}$.}
 \label{fig:1}
 \end{figure}
 %%%%%%%%%%%%%%%%%%%%%%%%%%%%%%%%%%%%%%%%%%%%%%%%%%%%%%%%%%%%%%%

The considered Josephson charge qubit is realized via a Cooper pair
box \cite{PRB_36_3548}: a nanometer-scale superconducting island,
which is connected via a Josephson junction to a large electrode
termed as a reservoir. The typical island dimensions is
1000nm$\times$50nm$ \times$20nm
(length$\times$width$\times$thickness) and the number of conduction
electrons is about $10^{7}-10^{8}$. If the superconducting energy
gap is large enough, it will effectively inhibit the particle tunnel
effect at low temperature, and it only allows Cooper-pair coherent
tunnel effect inside the superconducting Josephson junction. The two
symmetric dc SQUIDs are assumed to be identical and have same
Josephson coupling energy $E_{Jk}^{0}$ and same capacitance
$C_{Jk}$. Since the size of the loop is usually very small $(1\mu
m)$, the self-inductance effects of each SQUID loop can be ignored.
Each SQUID pierced by a magnetic flux $\Phi_{Xk}$ provides an
effective coupling energy given by $-E_{Jk}(\Phi_{Jk})\cos
\phi_{kA(B)}$ with $E_{Jk}(\Phi_{Jk})=2E_{Jk}^{0}
\cos(\pi\Phi_{Xk}/\Phi_{0}) $, where $\Phi_{0}=h/2e$ is the flux
quantum. The effective phase drop $\Phi_{kA(B)}$, with the subscript
$A(B)$ labeling the SQUID above (below) the island, equals the
average value $(\Phi_{kA(B)}^{L} +\Phi_{kA(B)}^{R})/2$ of the phase
that drops across the two Josephson junctions in the dc SQUID, where the
superscript $L(R)$ denotes the left (right) Josephson junction. The
phase drops $\phi_{kA}^{L}$ and $\phi_{kB}^{L}$ are related to the
total flux $\Phi=\Phi_{e}+LI$ through the inductance $L$ by the
constraint $\phi_{kB}^{L}- \phi_{kA}^{L}=2\pi\Phi/\Phi_{0}$, where
$\Phi_{e}$ is the externally applied magnetic flux threading the
inductance $L$.

In the muti-qubit circuit, we choose the bases $\{|0\rangle=
|n_{i}\rangle$, $|1\rangle=|n_{i}+1\rangle\}$ ($n_{i}$ is the number
of electrons in the $i$-th Cooper-pair box), then the Hamiltonian of
the system in the spin-1/2 representation can be reduced to
  \begin{eqnarray}
  \hat{H}=\varepsilon_{i}(V_{Xi})\sigma_{z}^{(i)}-
          \bar{E}_{Ji}(\Phi_{Xi},\Phi_{e},L)\sigma_{x}^{(i)},
  \end{eqnarray}
where $\sigma_{x,y,z}$ are Pauli matrices. $\varepsilon_{i}(V_{Xi})=
[{C_{i}V_{Xi}}/{e}-(2n_{i}+1)]E_{ci}/2$ is the charge energy that
can be controlled via the gate voltage, $E_{ci}=2{e^{2}}/(C_{i}+
C_{J0})$. The intrabit coupling $\bar{E}_{Ji}(\Phi_{Xi},
\Phi_{e},L)$ can be controlled by both the applied external flux
$\Phi_{e}$ through the common inductance, and the local flux
$\Phi_{Xi}$ through the two SQUID loops of the $i$-th Cooper-pair
box. According to Ref. \cite{PRL_89_197902}, $\bar{E}_{Ji}
\propto\cos(\pi\Phi_{e}/\Phi_{0})$, we choose $\Phi_{e}=\Phi_{0}/2$
for all boxes in Fig. 1, so that the intrabit coupling is
$\bar{E}_{Ji}=0$. We will discuss it in depth in the following text.

The inductance $L$ is shared by the Cooper-pair boxes $i$ and $j$ to
form the superconducting loops. The reduced Hamiltonian of the
system is given by
  \begin{eqnarray}
  \hat{H}=\sum_{k=i,j}[\varepsilon_{k}(V_{Xk})\sigma_{z}^{(k)}
  -\bar{E}_{Jk} \sigma_{x}^{(k)}]+J_{ij}\sigma_{x}^{(i)}\sigma_{x}^{(j)}.
  \end{eqnarray}
Here the interbit coupling $J_{ij}=-\pi^{2}LE_{Ji}E_{Jj}
\sin^{2}(\pi\Phi_{e}/$ $\Phi_{0})/\Phi_{0}^{2}$ is controlled by
both the external flux $\Phi_{e}$ and the local fluxes
$\Phi_{Xi,Xj}$. For simplicity, we switch $k=2$, so that the
Hamiltonian of the system is
  \begin{eqnarray}
  \hat{H}&=&[\varepsilon_{1}(V_{X1})\sigma_{z}^{(1)}-\bar{E}_{J1}\sigma_{x}^{(1)}]\nonumber \\
         &&+[\varepsilon_{2}(V_{X2})\sigma_{z}^{(2)}-\bar{E}_{J2}\sigma_{x}^{(2)}]
         +J_{12}\sigma_{x}^{(1)}\sigma_{x}^{(2)}.
  \end{eqnarray}
The state of the system at thermal equilibrium can be described by
the density operator $\rho=\texttt{exp}(-H/k_{B}T)/Z$, where
$Z=\texttt{tr}[\texttt{exp}(-H/k_BT)]$ is the partition function with
$T$ the temperature, and $k_{B}$ the Boltzmann's constant
\cite{PRA_66_044305}. We will discuss QD at finite temperatures;
this is called thermal QD.

We now propose the methods for achieving the possible maximum value
of QD:

\textbf{(i)} The intra-qubit coupling $\bar{E}_{Ji}=0$.

Our numerical results show QD does not exist if the
intra-qubit coupling $\bar{E}_{Ji} \neq 0$. Therefore, we only
consider the case $\bar{E}_{Ji}=0$ which can be achieved by
choosing $\Phi_{e}=\Phi_{0}/2$ for all boxes as $\bar{E}_{Ji}
(\Phi_{Xi}, \Phi_{e},L)=\xi_{ij} E_{Ji} \cos(\pi\Phi_{e}/\Phi_{0})$.
Then the Hamiltonian of the system reduces to $\hat{H}=
\varepsilon_{1} (V_{X1})\sigma_{z}^{(1)}+\varepsilon_{2}(V_{X2})
\sigma_{z}^{(2)} +J_{12}\sigma_{x}^{(1)}\sigma_{x}^{(2)}$, which is
of Ising-like \cite{PRB_60_11404} with the "magnetic field" along
the $z$ axis. For simplicity, we take $V_{X1}=V_{X2}=V_{X}$ and the
Hamiltonian of the system will be
  \begin{eqnarray}
  \hat{H}=\varepsilon(V_{X})(\sigma_{z}^{(1)}+\sigma_{z}^{(2)})+
  J_{12}\sigma_{x}^{(1)}\sigma_{x}^{(2)},
  \end{eqnarray}
the nonzero elements of the density operator $\rho$ are
  \begin{eqnarray}
  &&\rho_{11,44}=w_{\mp}/\alpha Z,~
  \rho_{22}=\rho_{33}=\cosh(\beta J)/Z,\nonumber\\
  &&\rho_{23}=\rho_{32}=-\sinh(\beta J)/Z,\nonumber\\
  &&\rho_{14}=\rho_{41}=-\gamma/\alpha Z,
  \end{eqnarray}
where $w_{\mp}=J_{12}^{2}[\lambda^{2}\cosh(\beta \lambda)\mp
2\varepsilon\lambda\sinh(\beta \lambda)]$,
$\alpha=J_{12}^4-12\varepsilon^4$,$\gamma=J_{12}^{3}
\lambda\sinh(\beta \lambda)$, $\lambda=\sqrt{4\varepsilon^{2}
+J_{12}^2}$, and $Z=2\cosh(\beta \lambda)+2\cosh(\beta J_{12})$.

\textbf{(ii)} The relationship between $J_{12}$ and $\varepsilon$.
  \begin{figure}
  \centering
  \resizebox{0.4\textwidth}{!}{%
  \includegraphics{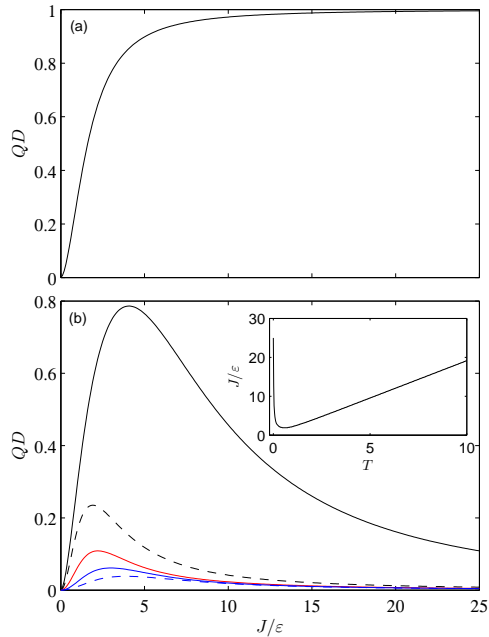}}
  \caption{QD versus $J_{12}/\varepsilon$. (a) shows the case of $k_BT=0$,
  while in (b) the lines from top to bottom correspond respectively to the
  cases of $k_BT=0.1K$, $0.5K$,
  $1K$, $1.5K$, and $2K$.
  The inset shows that the dependence of the critical $J/\varepsilon$ (when QD
  attains a certain maximum) on $T$. }
  \label{fig:2} \end{figure}
  %%%%%%%%%%%%%%%%%%%%%%%%%%%%%%%%%%%%%%%%%%%%%%%%%%%%%%%%%%

For $k_BT=0$, QD reaches its maximum value 1 asymptotically with the
increase of $J_{12}/\varepsilon$, see Fig. 2(a). For instance, QD is
about 0.9988 when $J_{12}/\varepsilon=25$. We display the
corresponding results in Fig. 2(b) for the nonzero temperature case,
from which one can see that the QD decreases with the increase of
$k_BT$. For any fixed nonzero $k_B T$, a certain maximum QD is
achieved at a critical $J_{12}/\varepsilon$ that depends on the
temperature. See the inset of Fig. 2(b), the critical
$J_{12}/\varepsilon$ decays with $T$ at the low temperature region,
and then $J_{12}/\varepsilon$ is discovered to be increased with the increase of $T$. This
phenomenon implies that the weak coupling is better for generating
the maximum QD in the low temperature region, while the case is
opposite in the high temperature region.

\subsection{equal magnetic flux $\Phi_{X1}=\Phi_{X2}$}

A small-size inductance can be made with Josephson junctions. By
fixing some system parameters one can see how the other parameters
affect variations of the thermal QD and EoF. In accordance with Ref.
\cite{PRL_89_197902}, we choose $L =30nH$. Since $J_{12}=
-(4\varepsilon_{0}^{2}\pi^{2}L/\Phi_{0}^{2}) \cos(\pi\Phi_{X1}/
\Phi_{0})$ $\cos(\pi\Phi_{X2}/\Phi_{0})$, $2k\pi\leq\Phi_{X1,2}/
\Phi_{0}\leq2(k+1)\pi$, $(k=0,1,2,...)$. We also chose the junction
capacitance $C_{J0}=10^{-5}F$, and used a small gate capacitance
$C=10^{-6}F$ to reduce the coupling of the environment. Using
$\varepsilon(V_{X})=[{CV_{X}}/{e}-(2n+1)] E_{c}/2$, $E_{c}=2{e^{2}}
/(C+C_{J0})$, we can get $V_{X}>10^{-6}V$ while QD
decreases with the increase of $V_{X}$.

   \begin{figure}
   \centering
   \resizebox{0.4\textwidth}{!}{%
   \includegraphics{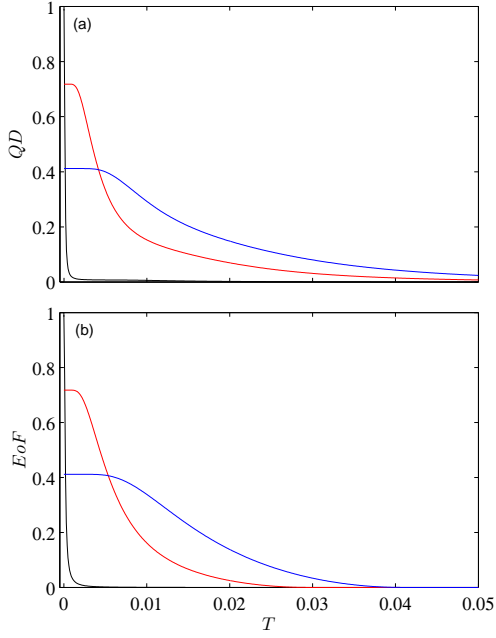}}
   \caption{(a) Temperature dependence of QD (a) and EoF (b), with $L=30nH$,
   $\Phi_{Xi}/\Phi_{0}=0$, and the black, red, and blue lines correspond
   to $V_{X}=7.5\mu V, 50\mu V$, and $100\mu V$, respectively.}
   \label{fig:3}
   \end{figure}

Fig. 3(a) is the dependence of QD on temperature $T$ with
$\Phi_{X1}=\Phi_{X2}=0$. Clearly, the QD decreases with the increase
of $T$ and this tendency is somewhat similar to that of EoF as shown
in Fig. 3(b). But the EoF diappears suddenly when $T$ reaches a
critical point, which is called entanglement sudden death (ESD)
\cite{Science_323_5914}. The critical temperature increases with the
increase of $V_X$. The reason for this behavior is due to the mixing
of the maximally entangled state with the other states, while at the
same time the thermal QD approaches asymptotically to zero if the
temperature is very high. The essence of this interesting phenomenon
is that the role of thermal fluctuations exceeds quantum cases as the
temperature grows. From Ollivier and Zurek's argument
\cite{PRL_88_017901}, we know that the absence of entanglement does
not mean classicality. The noisy environments can destroy the
quantumness of a system and degenerate it to a classical case
\cite{aop1}. QD measures total quantum correlations and it will not
disappear even with very high temperature. From this point one can
conclude that QD is more robust than entanglement \cite{aop1}.

If the thermal fluctuation is very strong, then the effects of
$V_{X},\ L$, and $\Phi_{Xi}/\Phi_{0}$ will become very weak. Now,
we will show how $V_{X}$ and $\Phi_{Xi}/\Phi_{0}$ affect the thermal QD
for the weak thermal fluctuation case with $L=30nH$. From Fig. 4
with $V_{X}=20\mu V$, one can see that both the thermal QD and the
entanglement behave as periodic functions of $\Phi_{Xi}/\Phi_{0}$.
The larger the value of $V_{X}$, the lower the amplitude of thermal
QD and EoF and they show the same periodic functions. This phenomena can be
interpreted by $J_{12}\propto \cos(\pi\Phi_{X1}/\Phi_{0})
\cos(\pi\Phi_{X2}/\Phi_{0})$, $\Phi_{Xi}/\Phi_{0}=\theta$. The QD
gets its maximum value at the ground state (the black line) when
$\Phi_{Xi}/\Phi_{0}=k~$ ($k\in\mathbb{N}$). For the thermal states,
the QD still presents periodic variations, but its maximum is reduced
when $\Phi_{Xi}/\Phi_{0}\neq k~$ ($k\in\mathbb{N}$). From the aforementioned example, one
can see that in order to achieve the needed QD, then the range of $\Phi_{Xi}/\Phi_{0}$ in one cycle is enough.

   \begin{figure}
   \centering
   \resizebox{0.4\textwidth}{!}{%
   \includegraphics{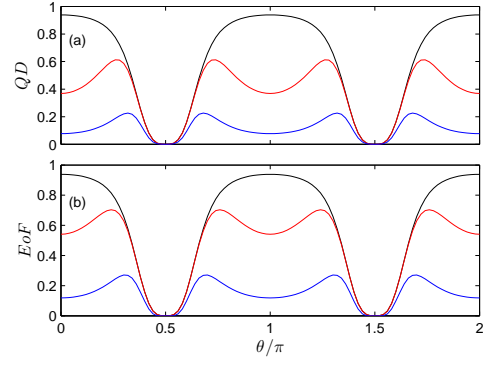}}
   \caption{QD (a) and EoF (b) versus $\Phi_{Xi}/\Phi_{0}=\theta$
   with $L=30nH$, $V_{Xk}=20\mu V$. The lines from top to bottom
   correspond to $T=0K$, $T=1\times 10^{-3}K$, and $T=5\times10^{-3}K$, respectively. } \label{fig:4}
   \end{figure}

\subsection{unequal magnetic flux $\Phi_{X1}\neq\Phi_{X2}$}

   \begin{figure}
   \centering \resizebox{0.4\textwidth}{!}{%
   \includegraphics{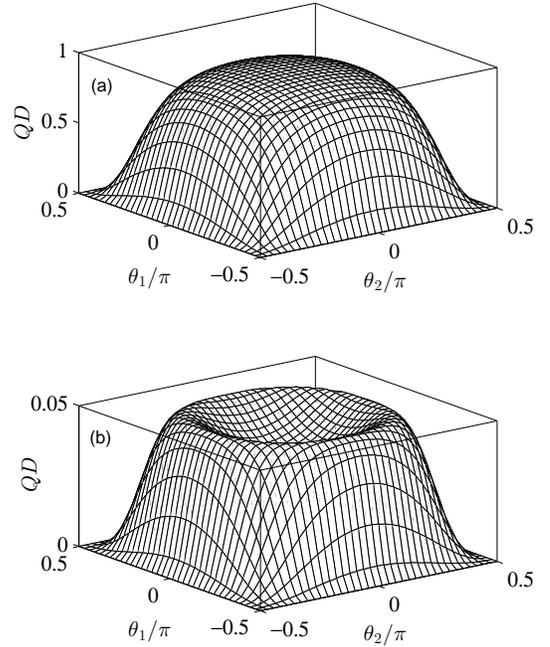}}
   \caption{QD versus $\theta_{1}=\Phi_{X1}/\Phi_{0}$
   and $\theta_{2}=\Phi_{X2}/\Phi_{0}$ with $L= 30nH$
   and $V_{X} = 20\mu V$ for $T=0$ (a)
   and $T=0.01K$ (b).}
   \label{fig:5}
   \end{figure}

In Section 3.1 we considered $\Phi_{X1}\doteq\Phi_{X2}$ and here we
consider $\Phi_{X1}\neq\Phi_{X2}$. For the ground state at
$k_{B}T=0$ and $V_{X}=20\mu V$, one can obtain the analytical
formulas
   \begin{eqnarray}
   D(\rho)=-u\log_{2}u-v\log_{2}v,
   \end{eqnarray}
where $u={(2\varepsilon+\lambda)^{2}}/{\zeta}$, $v={J^{2}}/{\zeta}$,
$\zeta=J^{2}+(2\varepsilon+\lambda)^{2}$. In Fig. 5(a), if
$\theta_{1}=\Phi_{X1}/\Phi_{0}$ and $\theta_{2}=\Phi_{X2}/\Phi_{0}$
change synchronously, when $k-\frac{1}{2}\leq\theta_{1,2}\leq k$
~($k\in\mathbb{N^+}$), QD increases with the increase of
$\theta_{1,2}$; when $k\leq\theta_{1,2}\leq k+\frac{1}{2}$
~($k\in\mathbb{N^+}$), QD decreases with the increase of
$\theta_{1,2}$. If $\theta_{1}$ and $\theta_{2}$ change
asynchronously when $k-\frac{1}{2}\leq\theta_{1}\leq k$
~($k\in\mathbb{N^+}$), QD decreases with the increase of
$\theta_{2}$ when $k\leq\theta_{2}\leq k-\frac{1}{2}$
~($k\in\mathbb{N^+}$); when $k\leq\theta_{1}\leq k+\frac{1}{2}$
~($k\in\mathbb{N^+}$), QD increases with the increase of
$\theta_{2}$ when $k-\frac{1}{2}\leq\theta_{2}\leq k$
~($k\in\mathbb{N^+}$). We explain this phenomenon by analyzing the
expression of QD in Eq. (8). When $V_{X}$ and $L$ are constant, QD
is only the function of $\Phi_{X1,2}$, which can be obtained from
$J_{12}=-(4\varepsilon_{0}^{2}\pi^{2}L/\Phi_{0}^{2})\cos(\pi\Phi_{X1}/
\Phi_{0})\cos(\pi\Phi_{X2}/\Phi_{0})$. For thermal states, the
curves in Fig. 5(b) and Fig. 4(a) have similar tendency, but QD is
very small when $T=0.01K$ and this is clearly not the desired result.
Thus, in order to obtain an ideal QD in the ground state one
should try to make $\Phi_{Xi}/\Phi_{0}=0$, while to obtain the
maximum QD in the thermal states one needs to adjust the system
parameters according to the actual situation.

\section{Conclusion}\label{sec:4}

In conclusion, one can make the values of QD as large as possible by
adjusting the parameters of our two qubits Josephson charge system.
For example, by taking $V_{Xk}= 20\mu V, L=30nH, \Phi_{X1,2}/
\Phi_{0}=k\pi, ~(k=0, 1, 2,...)$, the QD approaches approximately to
the maximum value 1 for the ground state case at $k_{B}T=0$.

Considering the effect of temperature $T$, thermal QD is more robust
than thermal entanglement. For example, thermal entanglement
undergoes sudden death while thermal QD does not. In theory, by
taking proper system parameters, one can always find a feasible
value of QD to help the experimenter to process the quantum
information. We hope our research findings demonstrated in
this paper will be experimentally realized in the future.

\begin{center}
\textbf{ACKNOWLEDGMENTS}
\end{center}

This work was supported by NSFC under Grant Nos. 11104217, 11174165
and 11275099. Shang thanks M.-L. Hu for his warmhearted discussion.
    \newcommand{\PRL}{Phys. Rev. Lett. }
    \newcommand{\RMP}{Rev. Mod. Phys. }
    \newcommand{\PRA}{Phys. Rev. A }
    \newcommand{\PRB}{Phys. Rev. B }
    \newcommand{\NJP}{New J. Phys. }
    \newcommand{\JPA}{J. Phys. A }
    \newcommand{\JPB}{J. Phys. B }
    \newcommand{\PLA}{Phys. Lett. A }
    \newcommand{\NP}{Nat. Phys. }
    \newcommand{\NC}{Nat. Commun. }
    \newcommand{\QIC}{Quantum Inf. Comput. }
    \newcommand{\QIP}{Quantum Inf. Process. }
    \newcommand{\EPJD}{Eur. Phys. J. D }
    %
    % BibTeX users please use
    % \bibliographystyle{}
    % \bibliography{}

\end{document}